\documentclass[reprint,aps,pra,twocolumn,showpacs,superscriptaddress]{revtex4-1}
\usepackage{graphicx}
\usepackage{amsmath}
\usepackage{amssymb}
\usepackage{epstopdf}
\usepackage{amsfonts}
\usepackage{dcolumn}
\usepackage{bm}

\usepackage{color}

\newcolumntype{d}[1]{D{.}{.}{#1}}

\begin{document}

\title{Universal and non-universal effective 
$N$-body interactions for ultracold
harmonically-trapped few-atom systems}
\author{X. Y. Yin}
\affiliation{Department of Physics and Astronomy, Washington State University, Pullman,
Washington 99164-2814, USA}
\author{D. Blume}
\affiliation{Department of Physics and Astronomy, Washington State University, Pullman,
Washington 99164-2814, USA}
\author{P. R. Johnson}
\affiliation{Department of Physics, American University, Washington DC 20016, USA}
\author{E. Tiesinga}
\affiliation{Joint Quantum Institute, National Institute of Standards and Technology \& University of Maryland, 100 Bureau Drive, Gaithersburg, Maryland 20899, USA}

\date{\today }

\begin{abstract}
We derive the ground-state energy for a small number of ultracold atoms in
an isotropic harmonic trap using effective quantum field theory (EFT). Atoms
are assumed to interact through pairwise energy-independent and
energy-dependent delta-function potentials with strengths proportional to
the scattering length $a$ and effective range volume $V$, respectively. The
calculations are performed systematically up to order $l^{-4}$, where $l$
denotes the harmonic oscillator length. The effective three-body interaction
contains a logarithmic divergence in the cutoff energy, giving rise to a
non-universal three-body interaction in the EFT. Our EFT results are
confirmed by nonperturbative numerical calculations for a Hamiltonian
with finite-range two-body Gaussian interactions. For this model Hamiltonian,
we explicitly calculate the non-universal effective three-body 
contribution to the energy.
\end{abstract}

\maketitle

\section{Introduction}
\label{sec_intro}
The properties of dilute Bose gases are to leading order determined by the
two-body free-space $s$-wave scattering length $a$~\cite{stringari_rmp}.
Two-body contact interactions between each pair of bosons are typically
assumed, and used to derive expansions around the non-interacting~\cite%
{yang57, yang57_2, huang59, luscher86, luscher91, wu59, braaten99,
braaten01, tan08, savage07, savage08, savage08_2, daily12, castin12, njp1,
njp2} or strongly-interacting unitary limit~\cite{kaplan98, jonsell02, 
braaten06, castin12, platter12, daily12}. 
Both few- and many-body systems have been
considered and, in some cases, the two limits have been connected using the
local density approximation \cite{stringari_rmp, lda1, gao03, jonsell02}. 
If the expansion is carried out to
sufficiently high order in $a$ or $1/a$, respectively, corrections due to
the two-body effective range volume $V$ have to be accounted for if a
consistent description that allows one to connect to atomic systems with
realistic interaction potentials is desired~\cite{yang57, huang59, tan08,
njp2, daily12, castin12_2, savage07, savage08, savage08_2}.

A question that has intrigued researchers for decades is how three-body
interactions come into play~\cite{wu59, efimov70, braaten97, braaten06,
witala02, tan08, savage07, savage08, savage08_2, petrov14}. In the
strongly-interacting regime, three-body physics manifests itself in the
Efimov effect. Signatures of the Efimov effect are seen by detecting atom
losses governed by the three-body recombination rate~\cite{braaten06,
greene99}. In contrast, we investigate in this work elastic three-body
scattering processes. We consider $N$ identical bosons with mass $M$ in a
spherically symmetric harmonic trap with angular frequency $\omega $ and
harmonic oscillator length $l=\sqrt{\hbar /(M\omega )}$ in the regime where
the two-body $s$-wave scattering length $a$ and two-body effective range
volume $V$ are small compared to the harmonic oscillator length $l$ and
volume $l^{3}$, respectively. The effective range volume $V$ is related to
the effective range $r_{\text{eff}}$ by 
\begin{equation}
V=\frac{1}{2}r_{\text{eff}}a^{2}.
\end{equation}
Earlier work developed a perturbative effective field theory (EFT) and
derived a low-energy Hamiltonian that accounts for terms up to order $%
(a/l)^{3}$ and 
$V/l^{3}$~\cite{njp2}. 
The resulting ground-state
energy was interpreted in terms of universal 
effective two-, three-, and four-body
interactions. The present paper extends this earlier work and determines
universal and non-universal
contributions of the terms proportional to $(a/l)^{4}$, $aV/l^{4}$, and $%
g_{3}^{(0)}/l^{4}$ to the ground-state energy; here, $g_{3}^{(0)}$ denotes a
three-body coupling constant.
Throughout this paper, the term universal is used to indicate
that the quantity under consideration is fully determined by the
low-energy two-body scattering observables. The term
non-universal, in contrast, is used to indicate that the quantity
under consideration cannot, in general, be determined from
the low-energy two-body scattering observables.

Our key findings are the following. ($i$) The $(a/l)^{4}$ term contains
effective five-, four-, three- and two-body interactions. The $aV/l^{4}$
term contains effective three- and two-body interactions. ($ii$) The
effective three-body interaction at order $l^{-4}$ contains a logarithmic
divergence in a cutoff energy $\Lambda $, introduced to regularize the EFT,
which signals a fundamental difference in character between the two- and
three-body interactions. Specifically, our results imply that the 
effective three-body
interaction contains a non-universal contribution 
that 
cannot be predicted from the low-energy two-body scattering observables.
Similar physics has previously been seen for the
homogeneous system~\cite{braaten99, braaten01, braaten02} and for few-body
systems confined to a periodic box~\cite{savage07, savage08, savage08_2,
tan08}. 
($iii$) We
extract
the non-universal three-body contribution 
from numerical ground state energies for $N=3-5$ bosons interacting via a
short-range two-body Gaussian model potential.

Section~\ref{sec_2} introduces the system Hamiltonian and summarizes our
final expression for the ground-state energy of the trapped $N$-boson
system. Sections~\ref{sec_3} and \ref{sec_4} discuss the structure of the terms at
order $l^{-4}$. In addition, Sec.~\ref{sec_4} elucidates that the field theoretical
treatment indicates the presence of a non-universal three-body interaction.
Lastly, Section~\ref{sec_5} summarizes our results and discusses implications.

\section{System Hamiltonian and ground-state energy}
\label{sec_2} 

We consider $N$ identical bosons with mass $M$ in a
three-dimensional isotropic harmonic trap with angular trapping frequency $%
\omega $. Our aim is to derive an expression for the ground-state energy of
the $N$-boson system, applicable in the low energy regime, using quantum
field theory~\cite{qft_book}. Our Hamiltonian is 
\begin{equation}
H=H_{1}+\sum_{p=2}^{N}\sum_{m=0,2,...}H_{p,\text{bare}}^{(m)},
\end{equation}
where $H_{1}$ denotes the single-particle Hamiltonian
\begin{equation}
H_{1}=\int \hat{\psi}^{\dagger }(\vec{r}_{1})\left( -\frac{\hbar ^{2}}{2M}%
\overrightarrow{\nabla }_{1}^{2}+\frac{1}{2}M\omega ^{2}{\vec{r}_{1}}%
^{2}\right) \hat{\psi}(\vec{r}_{1})\ d\vec{r}_{1}
\end{equation}%
and the bosonic field operators $\hat{\psi}(\vec{r})$ and $\hat{\psi}%
^{\dagger }(\vec{r})$ destroy and create particles at position $\vec{r},$
respectively. The term $H_{p,\text{bare}}^{(m)}$ denotes $p$-body contact
interactions
\begin{eqnarray}
H_{p,\text{bare}}^{(m)} &=&\frac{1}{p!}\int \hat{\psi}^{\dagger }(\vec{r}%
_{1})\cdots \hat{\psi}^{\dagger }(\vec{r}_{p})W_{p}^{(m)}(\vec{r}_{1},\vec{r}%
_{2},\cdots ,\vec{r}_{p})  \notag \\
&&~~~~\times \hat{\psi}(\vec{r}_{1})\cdots \hat{\psi}(\vec{r}_{p})\ d\vec{r}%
_{1}d\vec{r}_{2}\cdots d\vec{r}_{p}.
\end{eqnarray}
The superscript \textquotedblleft $(m)$\textquotedblright\ indicates the
order of the derivative operator in the $p$-body potentials $W_{p}^{(m)}$.
In our calculations, we expand the field operators in terms of the
eigenstates of the single-particle harmonic oscillator Hamiltonian~\cite%
{njp1, njp2}. 

Through order $l^{-4}$, we find that only three
potentials are needed: $W_{2}^{(0)}(\vec{r}_{1},\vec{r}_{2}),$ $W_{2}^{(2)}(%
\vec{r}_{1},\vec{r}_{2}),$ and $W_{3}^{(0)}(\vec{r}_{1},\vec{r}_{2},\vec{r}_{3});$ no
local four- or higher-body potentials are necessary. The two-body potential $%
W_{2}^{(0)}(\vec{r}_{1},\vec{r}_{2})$ corresponds to the \textquotedblleft
usual\textquotedblright\ $\delta $-function pseudopotential~\cite{fermi34}
\begin{equation}
W_{2}^{(0)}(\vec{r}_{1},\vec{r}_{2})=g_{2,\text{bare}}^{(0)}\delta (\vec{r}%
_{1}-\vec{r}_{2}),
\end{equation}%
where $g_{2,\text{bare}}^{(0)}$ is the two-body bare coupling constant. The $%
m=2$ two-body potential $W_{2}^{(2)}(\vec{r}_{1},\vec{r}_{2})$ depends on
the energy through the second-derivative operators~\cite{savage07, njp2}
\begin{eqnarray}
W_{2}^{(2)}(\vec{r}_{1},\vec{r}_{2}) &=&\frac{1}{2}g_{2,\text{bare}%
}^{(2)}\times   \notag  \label{eq_potential_2b_fr} \\
&&\left[ \overleftarrow{\nabla }_{12}^{2}\delta (\vec{r}_{1}-\vec{r}%
_{2})+\delta (\vec{r}_{1}-\vec{r}_{2})\overrightarrow{\nabla }_{12}^{2}%
\right] ,
\end{eqnarray}
where $g_{2,\text{bare}}^{(2)}$ is another two-body bare coupling constant.
The operators $\overleftarrow{\nabla }_{12}$ and $\overrightarrow{\nabla }%
_{12}$ are gradients with respect to the relative distance vector $\vec{r}%
_{1}-\vec{r}_{2}$ that act to the left and right, respectively. 
Note that the two-body interaction with $m=1$ is absent due to symmetry
constraints. The lowest order three-body potential is modeled by the product
of two $\delta $-functions, 
\begin{equation}
W_{3}^{(0)}(\vec{r}_{1},\vec{r}_{2},\vec{r}_{3})=g_{3,\text{bare}%
}^{(0)}\delta (\vec{r}_{1}-\vec{r}_{2})\delta (\vec{r}_{2}-\vec{r}_{3}),
\end{equation}%
where $g_{3,\text{bare}}^{(0)}$ is the three-body bare coupling constant.
The three-body potential acts only when three particles are at the same
position. 

We calculate the $N$-boson ground-state energy using renormalized
Rayleigh-Schr\"{o}dinger perturbation theory. Divergences arise at second-
and higher-order in perturbation theory~\cite{qft_book}. To obtain physical
results we include counterterm interactions for each $p$ and $m$
combination. Specifically, we write the bare coupling constant as~\cite%
{njp1, njp2} 
\begin{equation}
g_{p,\text{bare}}^{(m)}=g_{p}^{(m)}+g_{p,\text{ct}}^{(m)},
\end{equation}%
where $g_{p}^{(m)}$ is the \emph{physical} coupling constant and $g_{p,\text{%
ct}}^{(m)}$ the counterterm coupling constant. The counterterms $g_{2,\text{%
ct}}^{(0)}$ and $g_{2,\text{ct}}^{(2)}$ are determined self-consistently
such that the EFT energy shifts reproduce the ground-state energy for two
harmonically-trapped bosons interacting through a short-range potential with
free-space $s$-wave scattering length $a$ and free-space effective range
volume $V$ up to order $l^{-4}$ (see, e.g., Ref.~\cite{njp2}). In this
renormalization scheme, the physical coupling constants are 
\begin{equation}
g_{2}^{(0)}=\frac{4\pi \hbar ^{2}}{M}a\ \ \text{ and\ \  }g_{2}^{(2)}=-\frac{%
4\pi \hbar ^{2}}{M}V.
\end{equation}

We find it convenient to organize the contributions to the ground-state
energies $E_{N}$ in terms of powers of $1/l$~\cite{njp1, njp2}. 
To understand this
structure, it is instructive to perform a dimensional analysis. The coupling
constants $g_{p,\text{bare}}^{(m)}$, and correspondingly $g_{p}^{(m)}$, have
units of $energy\times (length)^{3p-3+m}$. For the scaled ground-state
energy $E_{N}/(\hbar \omega )$ this implies that the first-order correction
due to the Hamiltonian term proportional to $g_{2}^{(0)}$ corresponds to an
energy shift of order $1/l$. Similarly, the term proportional to
$(g_{2}^{(0)})^{2}$ corresponds to a shift of order $1/l^{2}$, and 
the terms proportional to
$(g_{2}^{(0)})^{3}$ and $g_{2}^{(2)}$
correspond to
shifts of order $1/l^{3}$.
Finally, the contributions $(g_{2}^{(0)})^{4}$, $g_{2}^{(0)}g_{2}^{(2)}$,
and $g_{3}^{(0)}$ lead to terms of order $1/l^{4}$. Thus, we can write the
scaled energy as%
\begin{equation}
\frac{E_{N}}{\hbar \omega }=\frac{3}{2}N+\sum_{p=2}^{N}\left( {N \atop p}
\right) U_{p}\,,  \label{eq_E}
\end{equation}%
where the dimensionless effective $p$-body interaction energies $U_{p}$ are
power series in $1/l$:
\begin{equation}
U_{p}=
\sum_{K=1}^{\infty}
\underbrace{\left[ \sum_{\substack{ %
k_{2,0},k_{2,2},k_{3,0} \\ k_{2,0}+3k_{2,2}+4k_{3,0}=K}}%
U_{p}^{(k_{2,0},k_{2,2},k_{3,0})}\right] }_{\mathcal{O}(l^{-K})}.
\end{equation}
The notation $\mathcal{O}(l^{-K})$ indicates that the term is proportional to
$l^{-K}$.
The dimensionless partial energies 
\begin{eqnarray}
\lefteqn{U_{p}^{(k_{2,0},k_{2,2},k_{3,0})}=c_{p}^{(k_{2,0},k_{2,2},k_{3,0})}}
\label{eq_U} \\
&&\times \left( \frac{g_{2}^{(0)}}{4\pi \hbar ^{2}/M}\frac{1}{l}\right)
^{k_{2,0}}\left( -\frac{g_{2}^{(2)}}{4\pi \hbar ^{2}/M}\frac{1}{l^{3}}%
\right) ^{k_{2,2}}\left( \frac{g_{3}^{(0)}}{\hbar ^{2}/M}\frac{1}{l^{4}}%
\right) ^{k_{3,0}}  \notag
\end{eqnarray}%
are proportional to $%
(g_{2}^{(0)})^{k_{2,0}}(g_{2}^{(2)})^{k_{2,2}}(g_{3}^{(0)})^{k_{3,0}}$. The
three superscripts $k_{p,m}$ take the values $0,1,2,\dots $ subject to the
constraint $k_{2,0}+3k_{2,2}+4k_{3,0}=K$; here, the prefactors of the $%
k_{p,m}$ are given by $3p+m-5$. The factors of $\pm 4\pi $ 
in the first two terms in the second line of
Eq.~(\ref{eq_U})
are included for later convenience.

Equation~(\ref{eq_E}) is valid when $a/l$, $V/l^{3}$, and $%
g_{3}^{(0)}/[(\hbar ^{2}/M)l^{4}]$ are much smaller than one. The
expansion coefficients $c_{p}^{(k_{2,0},k_{2,2},k_{3,0})}$ are summarized in
Table~\ref{table_ccoeff}. 
\begin{table*}[t]
\caption{Expansion coefficients $c_{p}^{(k_{2,0},k_{2,2},k_{3,0})}$, defined
in Eq.~(\protect\ref{eq_U}), of the effective $p$-body interactions for $N$
bosons in an isotropic harmonic trap, up to order $l^{-4}$. Columns 4 and 5
give analytic expressions and numerical values, respectively,
obtained using renormalized perturbation
theory. The numbers in round brackets in column 5 denote the
numerical uncertainty; those without error bars have been rounded. After
renormalization of the two-body interactions, all coefficients are finite
except for the logarithmically diverging $c_{3}^{(4,0,0)}.$ The terms $%
D_\text{a},D_\text{b},D_\text{c},$ and $D_\text{d}$ are defined in the text. 
To interpret Fig.~\ref{fig_1}(a), we calculate the $K=5$ effective
four-body contribution proportional to $Va^{2}$. This gives $%
c_{4}^{(2,1,0)}=25.42247.$ 
No other $K=5$ contributions are calculated in the present paper.
The function $\protect\zeta (z)$ is the Riemann Zeta
function. }\centering
\begin{tabular}{c c c c d{2.9}}
\hline
\hline
$p$ & $(k_{2,0},k_{2,2},k_{3,0})$ & $K$
& compact expression / comment 
&  \multicolumn{1}{c}{numerical value}\\
\hline
2 & $(1,0,0)$ & 1 & $(2/\pi)^{1/2}$ & 0.797884561 \\
  & $(2,0,0)$ & 2 & $(2/\pi)(1-\log2)$ & 0.195348572 \\
  & $(3,0,0)$ & 3 & $(2/\pi)^{3/2}(1-\frac{\pi^2}{24}
  -3\log2+\frac{3}{2}\log^{2}2)$ & -0.391118531 \\
  & $(0,1,0)$ & 3 & $(3/2)(2/\pi)^{1/2}$ & 1.196826841 \\
  & $(4,0,0)$ & 4 & $\frac{1}{3\pi^2}
                [12+\pi^2(-2+\log4)
                -4(-3+\log4)^2\log4
                -3\zeta(3)]$ 
                & -0.408766776 \\  
  & $(1,1,0)$ & 4 & $(1/\pi)(8-6\log2)$ & 1.222665489\\
\hline
3 & $(2,0,0)$ & 2 & $(2/\pi)[-4\sqrt{3}
                 +6-12\log2-6\log(2-\sqrt{3})]$ & -0.855758313 \\  
  & $(3,0,0)$ & 3 & see Ref.~\cite{njp2} 
  (sum evaluated numerically)& 2.7921(1)\\
  & $(4,0,0)$ & 4 & $10.8629(1)
             -12(2D_\text{a}
                +D_\text{b}
                +2D_\text{c}-3D_\text{d})$
                &  \multicolumn{1}{c}{log-divergent}\\
  & $(1,1,0)$ & 4 & $-(4/3\pi)[-36+26\sqrt{3}+9\log64
                 -27\log(2+\sqrt{3})]$
                   & -4.628397857 \\
  & $(0,0,1)$ & 4 & $16/(3\sqrt{3}\pi)$ & 0.980140259\\
\hline
4 & $(3,0,0)$ &  3 & see Ref.~\cite{njp2} 
(sum evaluated numerically)& 2.433174845\\
  & $(4,0,0)$ &  4  & 
  sum evaluated numerically  & -20.0(2)\\
\hline
5 & $(4,0,0)$ &  4 &
  sum evaluated numerically  & -11.12(2)\\
\hline
\hline
\end{tabular}
\label{table_ccoeff}
\end{table*}
After renormalization of the two-body interactions
all coefficients are finite except $c_{3}^{(4,0,0)}$, which diverges
logarithmically with the cutoff. The origin and implications of this
logarithmic divergence are discussed in Sec.~\ref{sec_4}. 
\begin{figure}[tbp]
\includegraphics[angle=0,width=80mm]{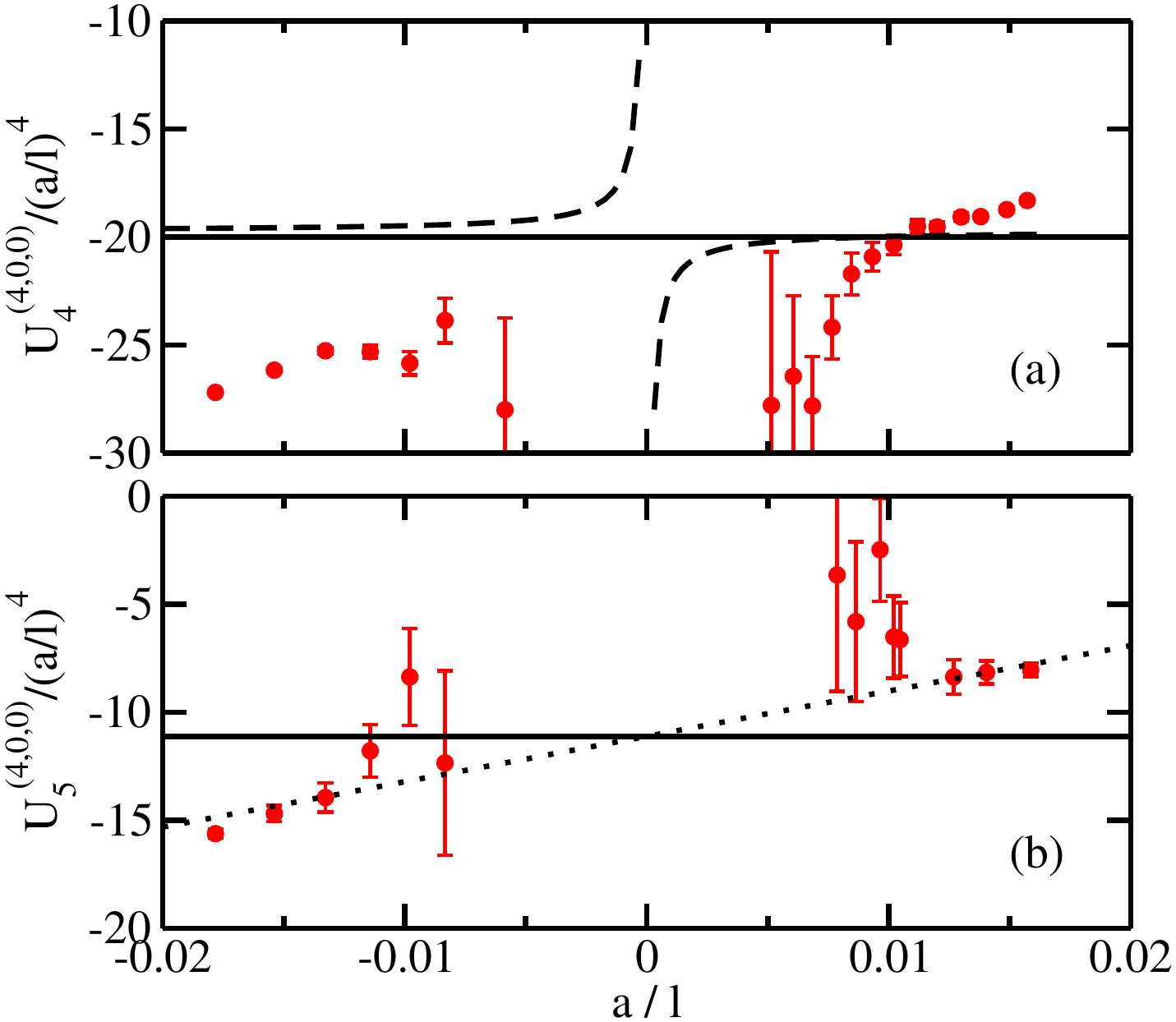}
\caption{(Color online) (a) and (b) show the effective four- and 
five-body contributions $U_{4}^{(4,0,0)}$ and $U_{5}^{(4,0,0)}$, respectively,
as a function of $a/l$. The interaction energies are scaled by 
$(a/l)^{-4},$ such that the EFT\ predictions at order $K=4$ are given by the
solid horizontal lines. Circles show numerical values for a 
model Hamiltonian with a pairwise
Gaussian interaction with $r_{0}=0.01l$. 
The numerical data is unreliable in the regime $|a/l|\lesssim0.007$
and $|a/l|\lesssim0.11$ for (a) and (b), respectively, as the
numerical uncertainty becomes comparable to or larger than $0.3$ times the
quantity of interest. The dashed line in panel (a) includes the scaled $K=5$
effective range volume dependent contribution, 
which is  proportional to $c_{4}^{(2,1,0)}V/a^{2}$. 
We use the numerically obtained effective range volume, 
as a function of $a,$ for
the Gaussian potential with $r_0=0.01l$, 
and $c_{4}^{(2,1,0)}=25.422472$ as determined within the EFT.
The dotted line in panel (b) shows a linear fit of the form $%
c_{5}^{(4,0,0)}+c_{5}^{(5,0,0)}a/l$ to the numerically determined energies
in the regime $|a/l|>0.11$,
with $c_{5}^{(4,0,0)}$ fixed at our EFT value of $-11.12$. We find $%
c_{4}^{(5,0,0)}\approx 210$.
The error bars, which are one standard deviation,
are estimated from the basis set extrapolation
errors of the numerically determined $N=3,4,$ and $5$ energies $E_N$.}
\label{fig_1}
\end{figure}
The $p=2$ coefficients
agree with what one obtains by expanding the exact zero-range solution for
two $s$-wave interacting particles in a harmonic 
trap~\cite{busch98, daily12}.

\section{The universal effective four- and five-body interactions}
\label{sec_3}

References~\cite{njp1, njp2} showed that the renormalized perturbation
theory treatment at orders $K=2$ and $3$ requires a counterterm coupling
constant $g_{2,\text{ct}}^{(0)}$, which cancels all divergences at these
orders. As we discuss now, new physics emerges at order $K=4$.

We start our discussion of the $K=4$ terms by considering the effective
four- and five-body interaction energies $U_{4}^{(4,0,0)}$ and 
$U_{5}^{(4,0,0)}$. The five-body term, which first arises at this order, is
finite. The four-body term is finite after renormalization of the two-body
interaction, with $g_{2,\text{ct}}^{(0)}$ removing power-law divergences.
Since $U_{4}^{(4,0,0)}$ and $U_{5}^{(4,0,0)}$ are fully determined
by $a/l$, we refer to these effective interactions as universal.
We were unable to evaluate the sums that give the coefficients $%
c_{4}^{(4,0,0)}$ and $c_{5}^{(4,0,0)}$ analytically. Numerical estimates and
uncertainties are reported in Table~\ref{table_ccoeff}.

To validate our EFT results for the effective
four- and five-body interactions, we compare
to numerical simulations of $N=2,3,4,$ and $5$ bosons interacting via a
finite-range, non-singular potential. We consider a Hamiltonian with
pairwise additive Gaussian model interaction $V_{\text{g}}(r)=V_{0}\exp
[-(r/r_{0})^{2}/2],$ with depth $V_{0}$ and width $r_{0}$, and determine the
energies $E_{N}$, $N>2$, numerically using an explicitly correlated Gaussian basis
set~\cite{njp2, cg_rmp, cg_book}. 
For $N=2$, we use a 
grid-based
B-spline approach.
For a given width $r_{0}$, we adjust the
depth $V_{0}$ ($V_{0}<0$ and $V_{0}>0$) such that $V_{\text{g}}(r)$
reproduces the desired physical free-space $s$-wave scattering length $a$ at
zero collision energy. The parameters are chosen such that $V_{\text{g}}(r)$
supports at most one bound state. 

The effective range volume for the Gaussian potential as a function of $V_{0}$
and thus scattering length $a$ was previously numerically calculated by us.
The result is shown in Fig.~3 of Ref.~\cite{njp2}. Crucial here is that in
the limit of zero scattering length the effective range volume approaches
zero. In fact, we have $V=-ar_{0}^{2}+\mathcal{O(}a^{2})$ from a
perturbative Born calculation of the two-body free-space scattering
amplitude.

Interestingly, following Refs.~\cite{savage07, savage08}, we can extract $%
U_{4}^{(4,0,0)}$ and $U_{5}^{(4,0,0)}$ from the numerically determined $E_{N}
$ using 
\begin{equation}
U_{4}^{(4,0,0)}=-U_{4}^{(3,0,0)}-6+(E_{4}-4E_{3}+6E_{2})/(\hbar \omega )+%
\mathcal{O}(l^{-5})
\end{equation}%
and 
\begin{equation}
U_{5}^{(4,0,0)}=15/2+(E_{5}-5E_{4}+10E_{3}-10E_{2})/(\hbar \omega )+%
\mathcal{O}(l^{-5}),
\end{equation}%
where the dimensionless partial energy 
$U_{4}^{(3,0,0)}$ has been obtained and validated in
Ref.~\cite{njp2}.

Figure~\ref{fig_1}(a) compares the numerically extracted scaled $%
U_{4}^{(4,0,0)}/(a/l)^4,$ for $r_{0}=0.01l,$ to the EFT prediction $%
c_{4}^{(4,0,0)}=-20.0$ given in Table~\ref{table_ccoeff}, as a function of $%
a/l.$ Similarly, Fig.~\ref{fig_1}(b) compares $U_{5}^{(4,0,0)}/(a/l)^4$ to
the EFT prediction $c_{5}^{(4,0,0)}=-11.12.$ In both cases, the EFT at order 
$K=4$ predicts horizontal lines. Comparison to the numerics shows reasonable
agreement, including the correct sign.

We can attempt to understand the deviations in Figs.~\ref{fig_1}(a)
and \ref{fig_1}(b) by looking at the 
$K=5$ contributions. The effective four-body interaction contains terms
proportional to $a^{5}$, $a^{2}V$, and $ag_{3}^{(0)}$. We have calculated the 
$a^{2}V$ coefficient from the EFT.
The dashed line in Fig.~\ref{fig_1}(a) shows the contribution proportional to
$c_4^{(2,1,0)}a^2V$, using the effective range volume $V$
for the Gaussian potential with $r_0=0.01l$. It can be seen that
this effective range volume correction to the solid line
is negligible in the regime for which our numerical data is reliable. Note
that as $V\propto -a$ for very small $|a|$, the correction diverges as 
$|a|\rightarrow0$. We conjecture that the deviation between the EFT
predictions for the effective four-body interaction
and the numerical data for $|a/l|\gtrsim0.01$ is due to both the $(a/l)^{5}$
and $ag_{3}^{(0)}$ contributions. Moreover we expect that $g_{3}^{(0)}$ depends
nontrivially on $a/l$ (see also Sec.~\ref{sec_4}).

The effective five-body interaction at order $K=5$ has only an $(a/l)^{5}$
contribution. As we have not calculated this contribution using EFT, the
numerical data in Fig.~\ref{fig_1}(b) is fit to a line
[see dotted line in Fig.~\ref{fig_1}(b)]
 with coefficients 
given in the caption of
Fig.~\ref{fig_1}. From the slope we 
extract an estimate for $%
c_{4}^{(5,0,0)}$.

\section{The non-universal three-body interaction}
\label{sec_4}

\begin{figure}[tbp]
\includegraphics[angle=0,width=75mm]{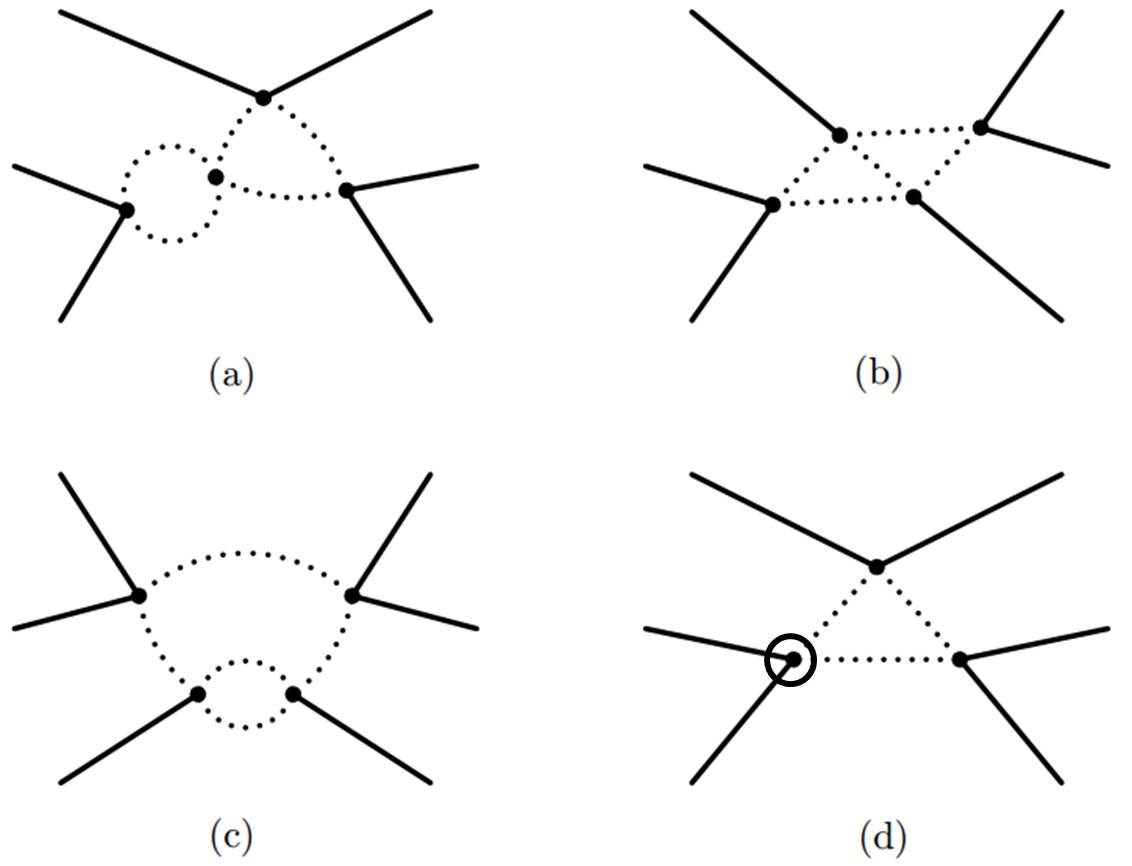}
\caption{Diagrammatic representation of the divergent sums that contribute
to the effective three-body interaction $U_{3}^{(4,0,0)}$. Diagrams (a)-(d)
represent the quantities $D_{\text{a}}$, $D_{\text{b}},D_{\text{c}},$ and $%
D_{\text{d}}$, (see Table~\protect\ref{table_ccoeff} and text). 
The dot represents
the two-body interaction with coupling constant $g_{2}^{(0)}$. The circled
dot represents the two-body counterterm with coupling constant $g_{2,\text{ct%
}}^{(0)}$.}
\label{fig_2}
\end{figure}

This section considers the effective three-body interaction.
Unlike
the four- and five-body terms, $U_{3}^{(4,0,0)}$ contains a logarithmic
divergence that is not renormalized by $g_{2,\text{ct}}^{(0)}$. To shed
light on this behavior, Figs.~\ref{fig_2}(a)-2(d) diagrammatically represent
the diverging sums $D_{\text{a}},$ $D_{\text{b}},$ $D_{\text{c}},$ and $D_{%
\text{d}}$ that enter into $U_{3}^{(4,0,0)}$. Note that these are modified
Rayleigh-Schr\"{o}dinger perturbation theory diagrams, using the formalism
described
in Refs.~\cite{njp1, njp2}, and not Feynman diagrams. 
For brevity, we do not
show the diagrams corresponding to convergent sums. Solid lines represent
particles in the single-particle ground state. Dotted lines represent
particles in single-particle excited states.
Vertices
represent interactions. The dot represents the two-body interaction with
coupling constant $g_{2}^{(0)},$ while the circled dot represents the
two-body counterterm with coupling constant $g_{2,\text{ct}}^{(0)}$. We
evaluate these diagrams numerically as a function of the cutoff energy $\Lambda ,
$ where terms corresponding to intermediate states with total energy greater
than $\Lambda $ are not included in the sums. We find that the diagrams
shown in Figs.~\ref{fig_2}(a), \ref{fig_2}(b), and \ref{fig_2}(d) diverge as 
$(\Lambda /\hbar \omega )^{1/2}$, $\log (\Lambda /\hbar \omega )$, and $%
(\Lambda /\hbar \omega )^{1/2}$, respectively. The diagram shown in Fig.~\ref%
{fig_2}(c) contains terms that diverge as $(\Lambda /\hbar \omega )^{1/2}$
and $\log (\Lambda /\hbar \omega )$.

\begin{figure}[tbp]
\includegraphics[angle=0,width=35mm]{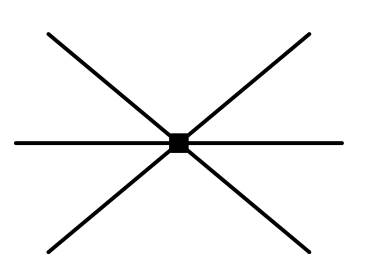}
\caption{Diagrammatic representation of the non-universal three-body
interaction. The square represents the three-body interaction with coupling
constant $g_{3}^{(0)}$. }
\label{fig_3}
\end{figure}

The power-law divergences contained in the diagrams
$D_\text{a}$ and $D_\text{c}$ [see Figs.~\ref{fig_2}(a) and %
\ref{fig_2}(c)] are renormalized by the two-body counterterm 
diagram $D_\text{d}$ [see Fig.~\ref{fig_2}(d)]. 
The $\log (\Lambda /\hbar \omega )$ divergences contained in the
diagrams $D_\text{b}$ and $D_\text{c}$ remain, however, and the
properly weighted diagrams $D_{\text{a}}$-$D_{\text{d}}$ evaluate to a term
of the form $q_{0}+q_{1}\log (\Lambda /\hbar \omega )$, where $q_{0}$ and $%
q_{1}$ are constants. This signals that a non-universal, local three-body
interaction with cutoff dependent coupling constant $g_{3}^{(0)},$
represented diagrammatically in Fig.~\ref{fig_3}, is needed~\cite{savage08,
braaten99, tan08}. 
Specifically, renormalization requires a three-body
interaction energy $U_{3}^{(0,0,1)},$ generated by $g_{3}^{(0)},$ which
cancels the logarithmic divergence in $U_{3}^{(4,0,0)}$. The corresponding $%
c_{3}^{(0,0,1)}$ value can be found in Table~\ref{table_ccoeff}.

The above discussion motivates us to define a
renormalization-scheme-independent three-body contribution (see also Refs.~%
\cite{savage07, savage08}) 
\begin{equation}
\bar{U}_{3}^{K=4}=U_{3}^{(4,0,0)}+U_{3}^{(0,0,1)}.
\end{equation}%
As $g_{3}^{(0)}$ is a new, undetermined parameter in the Hamiltonian, the
EFT does not make a unique prediction for $\bar{U}_{3}^{K=4}$ based on the
values of $g_{2}^{(0)}$ and $g_{2}^{(2)}.$ Instead, $\bar{U}_{3}^{K=4}$
depends on the short-range features of the true, \textquotedblleft
intrinsic\textquotedblright\ underlying interaction potentials. The
interaction energy $\bar{U}_{3}^{K=4}$ must therefore either be obtained by
measurement or by accurate numerical simulation 
of an $N$-body system ($N>2$).
We can extract the value of $\bar{U}_{3}^{K=4}$, to order $l^{-4}$,
using the numerically determined $N$-body ground state energies $E_N$,
\begin{eqnarray}
\label{eq_U3bar}
\binom{N}{3}\bar{U}_3^{K=4}=\frac{E_N}{\hbar\omega}
-\frac{3}{2}N  \nonumber\\
-\binom{N}{2}
\bigg[\sum_{k=1}^4 U_{2}^{(k,0,0)}+
U_{2}^{(0,1,0)}+U_{2}^{(1,1,0)}\bigg]  \nonumber\\
-\binom{N}{3}
\bigg[\sum_{k=2}^3 U_{3}^{(k,0,0)}
+U_{3}^{(1,1,0)}\bigg] \nonumber \\
-\binom{N}{4}
\bigg[\sum_{k=3}^4 U_{4}^{(k,0,0)}\bigg]
-\binom{N}{5}U_{5}^{(4,0,0)}
+\mathcal{O}\big(l^{-5}\big).
\end{eqnarray}
The key point is that the 
$U_{p}^{(k_{2,0},k_{2,2},k_{3,0})}$ quantities 
on the right hand side of Eq.~(\ref{eq_U3bar}) are known
from the EFT (see Table~\ref{table_ccoeff}).
This implies that we can calculate $\bar{U}_3^{K=4}$
for $N=3,4,5\cdots$,
provided the $E_N$ are known.

\begin{figure}[tbp]
\includegraphics[angle=0,width=80mm]{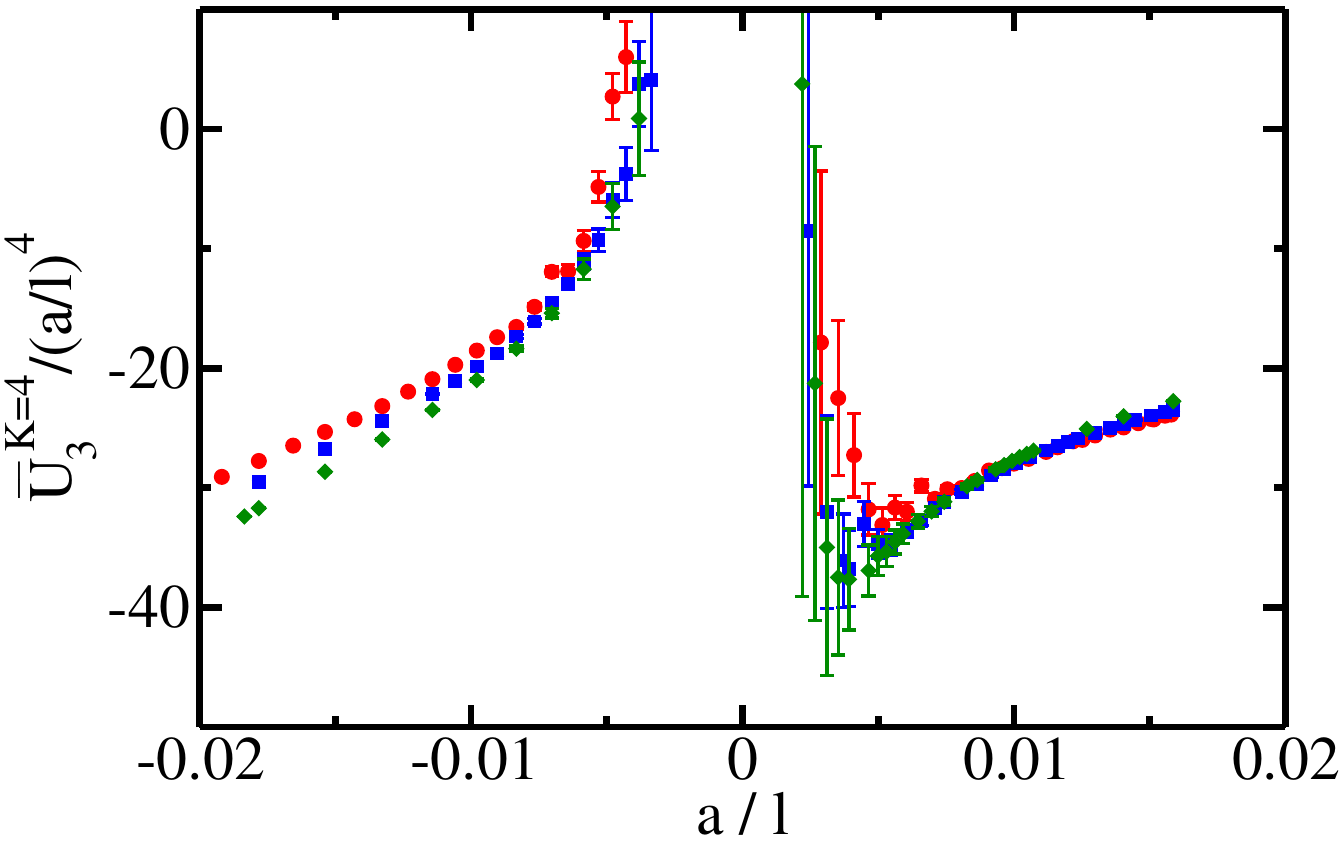}
\caption{(Color online) Scaled three-body interaction $\bar{U}
_{3}^{K=4}/(a/l)^{4}$ as a function of $a/l$, extracted from numerical
$N$-body ground-state energies for the Gaussian two-body
potential with width $r_{0}=0.01l$ and using Eq.~(\ref{eq_U3bar}). Circles, squares and
diamonds are determined from
Eq.~(\ref{eq_U3bar}) for $N=3,4,$ and $5$, respectively. The numerical data is unreliable
for $|a/l|\lesssim0.005,$ as the numerical uncertainty becomes comparable to or
larger than $0.3$ times the quantity of interest. 
The error bars, which are one standard deviation,
are estimated from the basis set extrapolation
errors of the numerically determined $N=3,4,$ and $5$ energies $E_N$.}
\label{fig_4}
\end{figure}

Figure~\ref{fig_4} shows $\bar{U}_{3}^{K=4}/(a/l)^{4}$ as a function of
$a/l$ determined from Eq.~(\ref{eq_U3bar}) for $N=3$, $4$, and $5$ using
the numerically determined ground-state energies for the Hamiltonian with
pairwise Gaussian interactions with width $r_{0}=0.01l$. 
We make two observations.  First, for a fixed
potential width $r_{0}$, the $\bar{U}_{3}^{K=4}$ calculated for $N=3$,
$4$, and $5$ collapse, to a good approximation, to a single curve. This
confirms that the extracted value of $\bar{U}_{3}^{K=4}$ scales with the
number of trimers in the system, i.e., that the physics seen is indeed
a three-body effect. Second, the fact that $\bar{U}_{3}^{K=4}/(a/l)^{4}$
is not independent of $a$ shows that $\bar{U}_{3}^{K=4} $ is not simply
proportional to $a^{4}$. This, combined with other analysis, indicates
that the three-body physics at order $K=4$ is not fully described by
the two-body $s$-wave scattering length and two-body effective range
volume.

To investigate the dependence of the non-universal three-body interaction on
the short-range interaction scale
of the Gaussian model interaction, 
we additionally calculated $\bar{U}%
_{3}^{K=4}/(a/l)^{4}$ for $r_{0}=0.005l$, $0.0075l$, and $0.0125l$. 
We find that the $\bar{U}_{3}^{K=4}$
for fixed $a$ but different $r_{0}$ differ
on the negative scattering length side where one expects the
formation of three-body bound states to be sensitive to the details of the
underlying two-body interaction model.
On the positive scattering length side, the $\bar{U}_{3}^{K=4}$
shows a comparatively weak dependence on $r_0$.
We believe that this can be attributed to the fact that the
purely repulsive Gaussian interaction model behaves 
similar to a hard core potential, especially for relatively
``large'' $a/l$. For the hardcore potential,
$\bar{U}_{3}^{K=4}$ has been shown to scale as $a^4$~\cite{tan08}.

If non-perturbative numerical $N$-body energies are not
available, we can still make rough, order-of-magnitude, estimates of
$\bar{U}_{3}^{K=4}$ by evaluating the
logarithmically diverging sums in the EFT up to the characteristic energy
scale of the two-body system, i.e., up to $\Lambda =\hbar ^{2}/(mr_{0}^{2})$%
. In practice, one might want to use the scale corresponding to the van der
Waals length as suggested in Ref.~\citep{braaten99}. In the present work,
however, it seems more appropriate to use the energy scale corresponding to
the Gaussian potential. For $r_{0}=0.01l$, this corresponds to $\Lambda
=10,000\hbar \omega $. Because we are unable to numerically evaluate the
necessary sums in diagrams $D_\text{a},D_\text{b},
D_\text{c}$ and $D_\text{d}$ of the EFT to a
value of $\Lambda $ this large, we instead extrapolate to $\Lambda
=10,000\hbar \omega $ using numerically determined estimates with smaller $%
\Lambda $ and the expected 
power-law and logarithmic divergences. This
approach yields $\bar{U}_{3}^{K=4}/(a/l)^{4}\approx -8.6$. Comparison with
Fig.~\ref{fig_4} shows that this estimate gives the correct sign and order
of magnitude of the true $\bar{U}_{3}^{K=4}$ for the Gaussian model
potential with $r_0=0.01l$.

Finally, we note that at order $K=4$, the effective three- and two-body
interactions $U_{3}^{(1,1,0)}$ and $U_{2}^{(1,1,0)}$ also depend on the
effective range volume. These universal contributions have been 
determined analytically and are given in Table~\ref{table_ccoeff}.
Following the convention introduced in Sec.~\ref{sec_intro},
we refer to
$U_{3}^{(1,1,0)}$ and $U_{2}^{(1,1,0)}$ 
as universal as they are fully determined by the low-energy two-body 
scattering properties, expressed in harmonic oscillator units.

\section{Conclusion}
\label{sec_5} 

In this paper, we utilized a quantum field theory approach to
derive the ground-state energy for a small number of bosons in a
spherically-symmetric harmonic trap  up to order $l^{-4}$, where $l$
is the harmonic oscillator length.  
We showed that the ground-state energy at this
order depends on two two-body parameters (the scattering length $a$ and
effective range volume $V$) and one emergent non-universal three-body
interaction strength ($g_{3}^{(0)}$). In the spirit of effective field
theory, these parameters can be determined by performing measurements at
two trap frequencies of the ground-state energy of the two-body system
and one measurement on the three-body system. Using these three data
points as input, the ground-state energy up to order $l^{-4} $ is then
known for any trapping frequency and any number of particles.

The emergence of the non-universal three-body interaction derived here for
harmonically-trapped atoms has been discussed for other systems. For
few-boson systems confined to a periodic box \cite{savage07, savage08,
savage08_2, tan08}, the ground-state energy can be organized, similar
to the harmonically-trapped system, in terms of powers of $1/L$
and $p$-body interactions, where $L$ is the length of the cubic
box. Interestingly, the leading order three-, four-, and five-body energy
contributions for $N$ bosons in the periodic box are proportional to
$a^{3}$, $a^{4}$, and $ a^{5}$, respectively~\cite{savage07, savage08,
savage08_2, tan08}, rather than the leading order contributions $a^{2}$,
$a^{3}$, and $a^{4}$ for bosons in a harmonic trap.  Just as for the
harmonically trapped system, the non-universal three-body interaction is
renormalization scheme dependent. Similar physics has also been observed
in the homogeneous system~\cite{braaten99}. Our analysis extends the
EFT approach to non-universal few-body interactions of harmonically
trapped bosons.

\section{Acknowledgements}

DB thanks Shina Tan and PRJ thanks Nathan Harshman for insightful
discussions. XYY and DB gratefully acknowledge support by the National
Science Foundation (NSF) through Grant No. PHY-1205443. PRJ and ET
acknowledge support from the US Army Research Office under contract/grant
60661PH. All four authors acknowledge support from the Institute for Nuclear
Theory during the program INT-14-1, \textquotedblleft Universality in
Few-Body Systems: Theoretical Challenges and New
Directions\textquotedblright .

\bibliography{mybib}

\begin{thebibliography}{35}%
\makeatletter
\providecommand \@ifxundefined [1]{%
 \@ifx{#1\undefined}
}%
\providecommand \@ifnum [1]{%
 \ifnum #1\expandafter \@firstoftwo
 \else \expandafter \@secondoftwo
 \fi
}%
\providecommand \@ifx [1]{%
 \ifx #1\expandafter \@firstoftwo
 \else \expandafter \@secondoftwo
 \fi
}%
\providecommand \natexlab [1]{#1}%
\providecommand \enquote  [1]{``#1''}%
\providecommand \bibnamefont  [1]{#1}%
\providecommand \bibfnamefont [1]{#1}%
\providecommand \citenamefont [1]{#1}%
\providecommand \href@noop [0]{\@secondoftwo}%
\providecommand \href [0]{\begingroup \@sanitize@url \@href}%
\providecommand \@href[1]{\@@startlink{#1}\@@href}%
\providecommand \@@href[1]{\endgroup#1\@@endlink}%
\providecommand \@sanitize@url [0]{\catcode `\\12\catcode `\$12\catcode
  `\&12\catcode `\#12\catcode `\^12\catcode `\_12\catcode `\%12\relax}%
\providecommand \@@startlink[1]{}%
\providecommand \@@endlink[0]{}%
\providecommand \url  [0]{\begingroup\@sanitize@url \@url }%
\providecommand \@url [1]{\endgroup\@href {#1}{\urlprefix }}%
\providecommand \urlprefix  [0]{URL }%
\providecommand \Eprint [0]{\href }%
\providecommand \doibase [0]{http://dx.doi.org/}%
\providecommand \selectlanguage [0]{\@gobble}%
\providecommand \bibinfo  [0]{\@secondoftwo}%
\providecommand \bibfield  [0]{\@secondoftwo}%
\providecommand \translation [1]{[#1]}%
\providecommand \BibitemOpen [0]{}%
\providecommand \bibitemStop [0]{}%
\providecommand \bibitemNoStop [0]{.\EOS\space}%
\providecommand \EOS [0]{\spacefactor3000\relax}%
\providecommand \BibitemShut  [1]{\csname bibitem#1\endcsname}%
\let\auto@bib@innerbib\@empty
%</preamble>
\bibitem [{\citenamefont {Dalfovo}\ \emph {et~al.}(1999)\citenamefont
  {Dalfovo}, \citenamefont {Giorgini}, \citenamefont {Pitaevskii},\ and\
  \citenamefont {Stringari}}]{stringari_rmp}%
  \BibitemOpen
  \bibfield  {author} {\bibinfo {author} {\bibfnamefont {F.}~\bibnamefont
  {Dalfovo}}, \bibinfo {author} {\bibfnamefont {S.}~\bibnamefont {Giorgini}},
  \bibinfo {author} {\bibfnamefont {L.~P.}\ \bibnamefont {Pitaevskii}}, \ and\
  \bibinfo {author} {\bibfnamefont {S.}~\bibnamefont {Stringari}},\ }\href
  {\doibase 10.1103/RevModPhys.71.463} {\bibfield  {journal} {\bibinfo
  {journal} {Rev. Mod. Phys.}\ }\textbf {\bibinfo {volume} {71}},\ \bibinfo
  {pages} {463} (\bibinfo {year} {1999})}\BibitemShut {NoStop}%
\bibitem [{\citenamefont {Huang}\ and\ \citenamefont {Yang}(1957)}]{yang57}%
  \BibitemOpen
  \bibfield  {author} {\bibinfo {author} {\bibfnamefont {K.}~\bibnamefont
  {Huang}}\ and\ \bibinfo {author} {\bibfnamefont {C.~N.}\ \bibnamefont
  {Yang}},\ }\href {\doibase 10.1103/PhysRev.105.767} {\bibfield  {journal}
  {\bibinfo  {journal} {Phys. Rev.}\ }\textbf {\bibinfo {volume} {105}},\
  \bibinfo {pages} {767} (\bibinfo {year} {1957})}\BibitemShut {NoStop}%
\bibitem [{\citenamefont {Lee}\ \emph {et~al.}(1957)\citenamefont {Lee},
  \citenamefont {Huang},\ and\ \citenamefont {Yang}}]{yang57_2}%
  \BibitemOpen
  \bibfield  {author} {\bibinfo {author} {\bibfnamefont {T.~D.}\ \bibnamefont
  {Lee}}, \bibinfo {author} {\bibfnamefont {K.}~\bibnamefont {Huang}}, \ and\
  \bibinfo {author} {\bibfnamefont {C.~N.}\ \bibnamefont {Yang}},\ }\href
  {\doibase 10.1103/PhysRev.106.1135} {\bibfield  {journal} {\bibinfo
  {journal} {Phys. Rev.}\ }\textbf {\bibinfo {volume} {106}},\ \bibinfo {pages}
  {1135} (\bibinfo {year} {1957})}\BibitemShut {NoStop}%
\bibitem [{\citenamefont {Huang}(1959)}]{huang59}%
  \BibitemOpen
  \bibfield  {author} {\bibinfo {author} {\bibfnamefont {K.}~\bibnamefont
  {Huang}},\ }\href {\doibase 10.1103/PhysRev.115.765} {\bibfield  {journal}
  {\bibinfo  {journal} {Phys. Rev.}\ }\textbf {\bibinfo {volume} {115}},\
  \bibinfo {pages} {765} (\bibinfo {year} {1959})}\BibitemShut {NoStop}%
\bibitem [{\citenamefont {L{\"{u}}scher}(1986)}]{luscher86}%
  \BibitemOpen
  \bibfield  {author} {\bibinfo {author} {\bibfnamefont {M.}~\bibnamefont
  {L{\"{u}}scher}},\ }\href {\doibase 10.1007/BF01211097} {\bibfield  {journal}
  {\bibinfo  {journal} {Commun. Math. Phys.}\ }\textbf {\bibinfo {volume}
  {105}},\ \bibinfo {pages} {153} (\bibinfo {year} {1986})}\BibitemShut
  {NoStop}%
\bibitem [{\citenamefont {L{\"{u}}scher}(1991)}]{luscher91}%
  \BibitemOpen
  \bibfield  {author} {\bibinfo {author} {\bibfnamefont {M.}~\bibnamefont
  {L{\"{u}}scher}},\ }\href {\doibase
  http://dx.doi.org/10.1016/0550-3213(91)90366-6} {\bibfield  {journal}
  {\bibinfo  {journal} {Nucl. Phys. B}\ }\textbf {\bibinfo {volume} {354}},\
  \bibinfo {pages} {531 } (\bibinfo {year} {1991})}\BibitemShut {NoStop}%
\bibitem [{\citenamefont {Wu}(1959)}]{wu59}%
  \BibitemOpen
  \bibfield  {author} {\bibinfo {author} {\bibfnamefont {T.~T.}\ \bibnamefont
  {Wu}},\ }\href {\doibase 10.1103/PhysRev.115.1390} {\bibfield  {journal}
  {\bibinfo  {journal} {Phys. Rev.}\ }\textbf {\bibinfo {volume} {115}},\
  \bibinfo {pages} {1390} (\bibinfo {year} {1959})}\BibitemShut {NoStop}%
\bibitem [{\citenamefont {Braaten}\ and\ \citenamefont
  {Nieto}(1999)}]{braaten99}%
  \BibitemOpen
  \bibfield  {author} {\bibinfo {author} {\bibfnamefont {E.}~\bibnamefont
  {Braaten}}\ and\ \bibinfo {author} {\bibfnamefont {A.}~\bibnamefont
  {Nieto}},\ }\href {\doibase 10.1007/s100510050925} {\bibfield  {journal}
  {\bibinfo  {journal} {EPJ B}\ }\textbf {\bibinfo {volume} {11}},\ \bibinfo
  {pages} {143} (\bibinfo {year} {1999})}\BibitemShut {NoStop}%
\bibitem [{\citenamefont {Braaten}\ \emph {et~al.}(2001)\citenamefont
  {Braaten}, \citenamefont {Hammer},\ and\ \citenamefont
  {Hermans}}]{braaten01}%
  \BibitemOpen
  \bibfield  {author} {\bibinfo {author} {\bibfnamefont {E.}~\bibnamefont
  {Braaten}}, \bibinfo {author} {\bibfnamefont {H.-W.}\ \bibnamefont {Hammer}},
  \ and\ \bibinfo {author} {\bibfnamefont {S.}~\bibnamefont {Hermans}},\ }\href
  {\doibase 10.1103/PhysRevA.63.063609} {\bibfield  {journal} {\bibinfo
  {journal} {Phys. Rev. A}\ }\textbf {\bibinfo {volume} {63}},\ \bibinfo
  {pages} {063609} (\bibinfo {year} {2001})}\BibitemShut {NoStop}%
\bibitem [{\citenamefont {Tan}(2008)}]{tan08}%
  \BibitemOpen
  \bibfield  {author} {\bibinfo {author} {\bibfnamefont {S.}~\bibnamefont
  {Tan}},\ }\href {\doibase 10.1103/PhysRevA.78.013636} {\bibfield  {journal}
  {\bibinfo  {journal} {Phys. Rev. A}\ }\textbf {\bibinfo {volume} {78}},\
  \bibinfo {pages} {013636} (\bibinfo {year} {2008})}\BibitemShut {NoStop}%
\bibitem [{\citenamefont {Beane}\ \emph {et~al.}(2007)\citenamefont {Beane},
  \citenamefont {Detmold},\ and\ \citenamefont {Savage}}]{savage07}%
  \BibitemOpen
  \bibfield  {author} {\bibinfo {author} {\bibfnamefont {S.~R.}\ \bibnamefont
  {Beane}}, \bibinfo {author} {\bibfnamefont {W.}~\bibnamefont {Detmold}}, \
  and\ \bibinfo {author} {\bibfnamefont {M.~J.}\ \bibnamefont {Savage}},\
  }\href {\doibase 10.1103/PhysRevD.76.074507} {\bibfield  {journal} {\bibinfo
  {journal} {Phys. Rev. D}\ }\textbf {\bibinfo {volume} {76}},\ \bibinfo
  {pages} {074507} (\bibinfo {year} {2007})}\BibitemShut {NoStop}%
\bibitem [{\citenamefont {Beane}\ \emph {et~al.}(2008)\citenamefont {Beane},
  \citenamefont {Detmold}, \citenamefont {Luu}, \citenamefont {Orginos},
  \citenamefont {Savage},\ and\ \citenamefont {Torok}}]{savage08}%
  \BibitemOpen
  \bibfield  {author} {\bibinfo {author} {\bibfnamefont {S.~R.}\ \bibnamefont
  {Beane}}, \bibinfo {author} {\bibfnamefont {W.}~\bibnamefont {Detmold}},
  \bibinfo {author} {\bibfnamefont {T.~C.}\ \bibnamefont {Luu}}, \bibinfo
  {author} {\bibfnamefont {K.}~\bibnamefont {Orginos}}, \bibinfo {author}
  {\bibfnamefont {M.~J.}\ \bibnamefont {Savage}}, \ and\ \bibinfo {author}
  {\bibfnamefont {A.}~\bibnamefont {Torok}},\ }\href {\doibase
  10.1103/PhysRevLett.100.082004} {\bibfield  {journal} {\bibinfo  {journal}
  {Phys. Rev. Lett.}\ }\textbf {\bibinfo {volume} {100}},\ \bibinfo {pages}
  {082004} (\bibinfo {year} {2008})}\BibitemShut {NoStop}%
\bibitem [{\citenamefont {Detmold}\ and\ \citenamefont
  {Savage}(2008)}]{savage08_2}%
  \BibitemOpen
  \bibfield  {author} {\bibinfo {author} {\bibfnamefont {W.}~\bibnamefont
  {Detmold}}\ and\ \bibinfo {author} {\bibfnamefont {M.~J.}\ \bibnamefont
  {Savage}},\ }\href {\doibase 10.1103/PhysRevD.77.057502} {\bibfield
  {journal} {\bibinfo  {journal} {Phys. Rev. D}\ }\textbf {\bibinfo {volume}
  {77}},\ \bibinfo {pages} {057502} (\bibinfo {year} {2008})}\BibitemShut
  {NoStop}%
\bibitem [{\citenamefont {Daily}\ \emph {et~al.}(2012)\citenamefont {Daily},
  \citenamefont {Yin},\ and\ \citenamefont {Blume}}]{daily12}%
  \BibitemOpen
  \bibfield  {author} {\bibinfo {author} {\bibfnamefont {K.~M.}\ \bibnamefont
  {Daily}}, \bibinfo {author} {\bibfnamefont {X.~Y.}\ \bibnamefont {Yin}}, \
  and\ \bibinfo {author} {\bibfnamefont {D.}~\bibnamefont {Blume}},\ }\href
  {\doibase 10.1103/PhysRevA.85.053614} {\bibfield  {journal} {\bibinfo
  {journal} {Phys. Rev. A}\ }\textbf {\bibinfo {volume} {85}},\ \bibinfo
  {pages} {053614} (\bibinfo {year} {2012})}\BibitemShut {NoStop}%
\bibitem [{\citenamefont {Werner}\ and\ \citenamefont
  {Castin}(2012{\natexlab{a}})}]{castin12}%
  \BibitemOpen
  \bibfield  {author} {\bibinfo {author} {\bibfnamefont {F.}~\bibnamefont
  {Werner}}\ and\ \bibinfo {author} {\bibfnamefont {Y.}~\bibnamefont
  {Castin}},\ }\href {\doibase 10.1103/PhysRevA.86.053633} {\bibfield
  {journal} {\bibinfo  {journal} {Phys. Rev. A}\ }\textbf {\bibinfo {volume}
  {86}},\ \bibinfo {pages} {053633} (\bibinfo {year}
  {2012}{\natexlab{a}})}\BibitemShut {NoStop}%
\bibitem [{\citenamefont {Johnson}\ \emph {et~al.}(2009)\citenamefont
  {Johnson}, \citenamefont {Tiesinga}, \citenamefont {Porto},\ and\
  \citenamefont {Williams}}]{njp1}%
  \BibitemOpen
  \bibfield  {author} {\bibinfo {author} {\bibfnamefont {P.~R.}\ \bibnamefont
  {Johnson}}, \bibinfo {author} {\bibfnamefont {E.}~\bibnamefont {Tiesinga}},
  \bibinfo {author} {\bibfnamefont {J.~V.}\ \bibnamefont {Porto}}, \ and\
  \bibinfo {author} {\bibfnamefont {C.~J.}\ \bibnamefont {Williams}},\ }\href
  {http://stacks.iop.org/1367-2630/11/i=9/a=093022} {\bibfield  {journal}
  {\bibinfo  {journal} {New J. Phys.}\ }\textbf {\bibinfo {volume} {11}},\
  \bibinfo {pages} {093022} (\bibinfo {year} {2009})}\BibitemShut {NoStop}%
\bibitem [{\citenamefont {Johnson}\ \emph {et~al.}(2012)\citenamefont
  {Johnson}, \citenamefont {Blume}, \citenamefont {Yin}, \citenamefont
  {Flynn},\ and\ \citenamefont {Tiesinga}}]{njp2}%
  \BibitemOpen
  \bibfield  {author} {\bibinfo {author} {\bibfnamefont {P.~R.}\ \bibnamefont
  {Johnson}}, \bibinfo {author} {\bibfnamefont {D.}~\bibnamefont {Blume}},
  \bibinfo {author} {\bibfnamefont {X.~Y.}\ \bibnamefont {Yin}}, \bibinfo
  {author} {\bibfnamefont {W.~F.}\ \bibnamefont {Flynn}}, \ and\ \bibinfo
  {author} {\bibfnamefont {E.}~\bibnamefont {Tiesinga}},\ }\href
  {http://stacks.iop.org/1367-2630/14/i=5/a=053037} {\bibfield  {journal}
  {\bibinfo  {journal} {New J. Phys.}\ }\textbf {\bibinfo {volume} {14}},\
  \bibinfo {pages} {053037} (\bibinfo {year} {2012})}\BibitemShut {NoStop}%
\bibitem [{\citenamefont {Kaplan}\ \emph {et~al.}(1998)\citenamefont {Kaplan},
  \citenamefont {Savage},\ and\ \citenamefont {Wise}}]{kaplan98}%
  \BibitemOpen
  \bibfield  {author} {\bibinfo {author} {\bibfnamefont {D.~B.}\ \bibnamefont
  {Kaplan}}, \bibinfo {author} {\bibfnamefont {M.~J.}\ \bibnamefont {Savage}},
  \ and\ \bibinfo {author} {\bibfnamefont {M.~B.}\ \bibnamefont {Wise}},\
  }\href {\doibase http://dx.doi.org/10.1016/S0370-2693(98)00210-X} {\bibfield
  {journal} {\bibinfo  {journal} {Phys. Lett. B}\ }\textbf {\bibinfo {volume}
  {424}},\ \bibinfo {pages} {390 } (\bibinfo {year} {1998})}\BibitemShut
  {NoStop}%
\bibitem [{\citenamefont {Jonsell}\ \emph {et~al.}(2002)\citenamefont
  {Jonsell}, \citenamefont {Heiselberg},\ and\ \citenamefont
  {Pethick}}]{jonsell02}%
  \BibitemOpen
  \bibfield  {author} {\bibinfo {author} {\bibfnamefont {S.}~\bibnamefont
  {Jonsell}}, \bibinfo {author} {\bibfnamefont {H.}~\bibnamefont {Heiselberg}},
  \ and\ \bibinfo {author} {\bibfnamefont {C.~J.}\ \bibnamefont {Pethick}},\
  }\href {\doibase 10.1103/PhysRevLett.89.250401} {\bibfield  {journal}
  {\bibinfo  {journal} {Phys. Rev. Lett.}\ }\textbf {\bibinfo {volume} {89}},\
  \bibinfo {pages} {250401} (\bibinfo {year} {2002})}\BibitemShut {NoStop}%
\bibitem [{\citenamefont {Braaten}\ and\ \citenamefont
  {Hammer}(2006)}]{braaten06}%
  \BibitemOpen
  \bibfield  {author} {\bibinfo {author} {\bibfnamefont {E.}~\bibnamefont
  {Braaten}}\ and\ \bibinfo {author} {\bibfnamefont {H.-W.}\ \bibnamefont
  {Hammer}},\ }\href {\doibase http://dx.doi.org/10.1016/j.physrep.2006.03.001}
  {\bibfield  {journal} {\bibinfo  {journal} {Phys. Rep.}\ }\textbf {\bibinfo
  {volume} {428}},\ \bibinfo {pages} {259 } (\bibinfo {year}
  {2006})}\BibitemShut {NoStop}%
\bibitem [{\citenamefont {Ji}\ \emph {et~al.}(2012)\citenamefont {Ji},
  \citenamefont {Phillips},\ and\ \citenamefont {Platter}}]{platter12}%
  \BibitemOpen
  \bibfield  {author} {\bibinfo {author} {\bibfnamefont {C.}~\bibnamefont
  {Ji}}, \bibinfo {author} {\bibfnamefont {D.~R.}\ \bibnamefont {Phillips}}, \
  and\ \bibinfo {author} {\bibfnamefont {L.}~\bibnamefont {Platter}},\ }\href
  {\doibase http://dx.doi.org/10.1016/j.aop.2012.02.001} {\bibfield  {journal}
  {\bibinfo  {journal} {Annals Phys.}\ }\textbf {\bibinfo {volume} {327}},\
  \bibinfo {pages} {1803 } (\bibinfo {year} {2012})}\BibitemShut {NoStop}%
\bibitem [{\citenamefont {Fabrocini}\ and\ \citenamefont {Polls}(1999)}]{lda1}%
  \BibitemOpen
  \bibfield  {author} {\bibinfo {author} {\bibfnamefont {A.}~\bibnamefont
  {Fabrocini}}\ and\ \bibinfo {author} {\bibfnamefont {A.}~\bibnamefont
  {Polls}},\ }\href {\doibase 10.1103/PhysRevA.60.2319} {\bibfield  {journal}
  {\bibinfo  {journal} {Phys. Rev. A}\ }\textbf {\bibinfo {volume} {60}},\
  \bibinfo {pages} {2319} (\bibinfo {year} {1999})}\BibitemShut {NoStop}%
\bibitem [{\citenamefont {Fu}\ \emph {et~al.}(2003)\citenamefont {Fu},
  \citenamefont {Wang},\ and\ \citenamefont {Gao}}]{gao03}%
  \BibitemOpen
  \bibfield  {author} {\bibinfo {author} {\bibfnamefont {H.}~\bibnamefont
  {Fu}}, \bibinfo {author} {\bibfnamefont {Y.}~\bibnamefont {Wang}}, \ and\
  \bibinfo {author} {\bibfnamefont {B.}~\bibnamefont {Gao}},\ }\href {\doibase
  10.1103/PhysRevA.67.053612} {\bibfield  {journal} {\bibinfo  {journal} {Phys.
  Rev. A}\ }\textbf {\bibinfo {volume} {67}},\ \bibinfo {pages} {053612}
  (\bibinfo {year} {2003})}\BibitemShut {NoStop}%
\bibitem [{\citenamefont {Werner}\ and\ \citenamefont
  {Castin}(2012{\natexlab{b}})}]{castin12_2}%
  \BibitemOpen
  \bibfield  {author} {\bibinfo {author} {\bibfnamefont {F.}~\bibnamefont
  {Werner}}\ and\ \bibinfo {author} {\bibfnamefont {Y.}~\bibnamefont
  {Castin}},\ }\href {\doibase 10.1103/PhysRevA.86.013626} {\bibfield
  {journal} {\bibinfo  {journal} {Phys. Rev. A}\ }\textbf {\bibinfo {volume}
  {86}},\ \bibinfo {pages} {013626} (\bibinfo {year}
  {2012}{\natexlab{b}})}\BibitemShut {NoStop}%
\bibitem [{\citenamefont {Efimov}(1970)}]{efimov70}%
  \BibitemOpen
  \bibfield  {author} {\bibinfo {author} {\bibfnamefont {V.}~\bibnamefont
  {Efimov}},\ }\href {\doibase http://dx.doi.org/10.1016/0370-2693(70)90349-7}
  {\bibfield  {journal} {\bibinfo  {journal} {Phys. Lett. B}\ }\textbf
  {\bibinfo {volume} {33}},\ \bibinfo {pages} {563 } (\bibinfo {year}
  {1970})}\BibitemShut {NoStop}%
\bibitem [{\citenamefont {Braaten}\ and\ \citenamefont
  {Nieto}(1997)}]{braaten97}%
  \BibitemOpen
  \bibfield  {author} {\bibinfo {author} {\bibfnamefont {E.}~\bibnamefont
  {Braaten}}\ and\ \bibinfo {author} {\bibfnamefont {A.}~\bibnamefont
  {Nieto}},\ }\href {\doibase 10.1103/PhysRevB.55.8090} {\bibfield  {journal}
  {\bibinfo  {journal} {Phys. Rev. B}\ }\textbf {\bibinfo {volume} {55}},\
  \bibinfo {pages} {8090} (\bibinfo {year} {1997})}\BibitemShut {NoStop}%
\bibitem [{\citenamefont {Epelbaum}\ \emph {et~al.}(2002)\citenamefont
  {Epelbaum}, \citenamefont {Nogga}, \citenamefont {Gl\"ockle}, \citenamefont
  {Kamada}, \citenamefont {Mei\ss{}ner},\ and\ \citenamefont
  {Wita\l{}a}}]{witala02}%
  \BibitemOpen
  \bibfield  {author} {\bibinfo {author} {\bibfnamefont {E.}~\bibnamefont
  {Epelbaum}}, \bibinfo {author} {\bibfnamefont {A.}~\bibnamefont {Nogga}},
  \bibinfo {author} {\bibfnamefont {W.}~\bibnamefont {Gl\"ockle}}, \bibinfo
  {author} {\bibfnamefont {H.}~\bibnamefont {Kamada}}, \bibinfo {author}
  {\bibfnamefont {U.-G.}\ \bibnamefont {Mei\ss{}ner}}, \ and\ \bibinfo {author}
  {\bibfnamefont {H.}~\bibnamefont {Wita\l{}a}},\ }\href {\doibase
  10.1103/PhysRevC.66.064001} {\bibfield  {journal} {\bibinfo  {journal} {Phys.
  Rev. C}\ }\textbf {\bibinfo {volume} {66}},\ \bibinfo {pages} {064001}
  (\bibinfo {year} {2002})}\BibitemShut {NoStop}%
\bibitem [{\citenamefont {Petrov}(2014)}]{petrov14}%
  \BibitemOpen
  \bibfield  {author} {\bibinfo {author} {\bibfnamefont {D.~S.}\ \bibnamefont
  {Petrov}},\ }\href {\doibase 10.1103/PhysRevLett.112.103201} {\bibfield
  {journal} {\bibinfo  {journal} {Phys. Rev. Lett.}\ }\textbf {\bibinfo
  {volume} {112}},\ \bibinfo {pages} {103201} (\bibinfo {year}
  {2014})}\BibitemShut {NoStop}%
\bibitem [{\citenamefont {Esry}\ \emph {et~al.}(1999)\citenamefont {Esry},
  \citenamefont {Greene},\ and\ \citenamefont {Burke}}]{greene99}%
  \BibitemOpen
  \bibfield  {author} {\bibinfo {author} {\bibfnamefont {B.~D.}\ \bibnamefont
  {Esry}}, \bibinfo {author} {\bibfnamefont {C.~H.}\ \bibnamefont {Greene}}, \
  and\ \bibinfo {author} {\bibfnamefont {J.~P.}\ \bibnamefont {Burke}},\ }\href
  {\doibase 10.1103/PhysRevLett.83.1751} {\bibfield  {journal} {\bibinfo
  {journal} {Phys. Rev. Lett.}\ }\textbf {\bibinfo {volume} {83}},\ \bibinfo
  {pages} {1751} (\bibinfo {year} {1999})}\BibitemShut {NoStop}%
\bibitem [{\citenamefont {Braaten}\ \emph {et~al.}(2002)\citenamefont
  {Braaten}, \citenamefont {Hammer},\ and\ \citenamefont {Mehen}}]{braaten02}%
  \BibitemOpen
  \bibfield  {author} {\bibinfo {author} {\bibfnamefont {E.}~\bibnamefont
  {Braaten}}, \bibinfo {author} {\bibfnamefont {H.-W.}\ \bibnamefont {Hammer}},
  \ and\ \bibinfo {author} {\bibfnamefont {T.}~\bibnamefont {Mehen}},\ }\href
  {\doibase 10.1103/PhysRevLett.88.040401} {\bibfield  {journal} {\bibinfo
  {journal} {Phys. Rev. Lett.}\ }\textbf {\bibinfo {volume} {88}},\ \bibinfo
  {pages} {040401} (\bibinfo {year} {2002})}\BibitemShut {NoStop}%
\bibitem [{\citenamefont {Srednicki}(2007)}]{qft_book}%
  \BibitemOpen
  \bibfield  {author} {\bibinfo {author} {\bibfnamefont {M.}~\bibnamefont
  {Srednicki}},\ }\href {http://books.google.com/books?id=5OepxIG42B4C} {\emph
  {\bibinfo {title} {Quantum Field Theory}}}\ (\bibinfo  {publisher} {Cambridge
  University Press},\ \bibinfo {year} {2007})\BibitemShut {NoStop}%
\bibitem [{\citenamefont {Fermi}(1934)}]{fermi34}%
  \BibitemOpen
  \bibfield  {author} {\bibinfo {author} {\bibfnamefont {E.}~\bibnamefont
  {Fermi}},\ }\href@noop {} {\bibfield  {journal} {\bibinfo  {journal} {Nuovo
  Cimento}\ }\textbf {\bibinfo {volume} {11}},\ \bibinfo {pages} {157}
  (\bibinfo {year} {1934})}\BibitemShut {NoStop}%
\bibitem [{\citenamefont {Busch}\ \emph {et~al.}(1998)\citenamefont {Busch},
  \citenamefont {Englert}, \citenamefont {Rzażewski},\ and\ \citenamefont
  {Wilkens}}]{busch98}%
  \BibitemOpen
  \bibfield  {author} {\bibinfo {author} {\bibfnamefont {T.}~\bibnamefont
  {Busch}}, \bibinfo {author} {\bibfnamefont {B.-G.}\ \bibnamefont {Englert}},
  \bibinfo {author} {\bibfnamefont {K.}~\bibnamefont {Rzażewski}}, \ and\
  \bibinfo {author} {\bibfnamefont {M.}~\bibnamefont {Wilkens}},\ }\href
  {\doibase 10.1023/A:1018705520999} {\bibfield  {journal} {\bibinfo  {journal}
  {Found. of Phys.}\ }\textbf {\bibinfo {volume} {28}},\ \bibinfo {pages} {549}
  (\bibinfo {year} {1998})}\BibitemShut {NoStop}%
\bibitem [{\citenamefont {Mitroy}\ \emph {et~al.}(2013)\citenamefont {Mitroy},
  \citenamefont {Bubin}, \citenamefont {Horiuchi}, \citenamefont {Suzuki},
  \citenamefont {Adamowicz}, \citenamefont {Cencek}, \citenamefont {Szalewicz},
  \citenamefont {Komasa}, \citenamefont {Blume},\ and\ \citenamefont
  {Varga}}]{cg_rmp}%
  \BibitemOpen
  \bibfield  {author} {\bibinfo {author} {\bibfnamefont {J.}~\bibnamefont
  {Mitroy}}, \bibinfo {author} {\bibfnamefont {S.}~\bibnamefont {Bubin}},
  \bibinfo {author} {\bibfnamefont {W.}~\bibnamefont {Horiuchi}}, \bibinfo
  {author} {\bibfnamefont {Y.}~\bibnamefont {Suzuki}}, \bibinfo {author}
  {\bibfnamefont {L.}~\bibnamefont {Adamowicz}}, \bibinfo {author}
  {\bibfnamefont {W.}~\bibnamefont {Cencek}}, \bibinfo {author} {\bibfnamefont
  {K.}~\bibnamefont {Szalewicz}}, \bibinfo {author} {\bibfnamefont
  {J.}~\bibnamefont {Komasa}}, \bibinfo {author} {\bibfnamefont
  {D.}~\bibnamefont {Blume}}, \ and\ \bibinfo {author} {\bibfnamefont
  {K.}~\bibnamefont {Varga}},\ }\href {\doibase 10.1103/RevModPhys.85.693}
  {\bibfield  {journal} {\bibinfo  {journal} {Rev. Mod. Phys.}\ }\textbf
  {\bibinfo {volume} {85}},\ \bibinfo {pages} {693} (\bibinfo {year}
  {2013})}\BibitemShut {NoStop}%
\bibitem [{\citenamefont {Suzuki}\ and\ \citenamefont {Varga}(1998)}]{cg_book}%
  \BibitemOpen
  \bibfield  {author} {\bibinfo {author} {\bibfnamefont {Y.}~\bibnamefont
  {Suzuki}}\ and\ \bibinfo {author} {\bibfnamefont {K.}~\bibnamefont {Varga}},\
  }\href {http://books.google.com/books?id=BpyTpm1VLx8C} {\emph {\bibinfo
  {title} {Stochastic Variational Approach to Quantum-Mechanical Few-Body
  Problems}}}\ (\bibinfo  {publisher} {Springer},\ \bibinfo {year}
  {1998})\BibitemShut {NoStop}%
\end{thebibliography}%

\end{document}